\documentclass{aa}
\usepackage{graphicx}
\usepackage{txfonts}
\begin{document}
   \titlerunning{The CSE of IRC+10216 from milli-arcsecond to arcmin scales}
   \title{The circumstellar envelope of IRC+10216 from milli-arcsecond to arcmin scales
   \thanks{Based on observations collected with the VLT/Antu and Yepun telescopes
          (Paranal Observatory, ESO, Chile) using the FORS1 and NACO instruments
          (programs ID 63.I-0177A, 70.C-0565A, 70.C-0271B, 70.D-0271B).}}

   
   \author{I.C. Le\~ao\inst{1,2} \and P. de Laverny\inst{1} \and D. M\'ekarnia\inst{1}
           \and J.R. De Medeiros\inst{2} \and B. Vandame\inst{3}}
   
   \institute{Observatoire de la C\^ote d'Azur,
              Dpt Cassiop\'ee, CNRS - UMR 6202, BP 4229, 06304 Nice Cedex 4, France\\
              \email{[leao;laverny;mekarnia]@obs-nice.fr}
         \and Departamento de F\'isica, Universidade Federal do Rio Grande do Norte,
              59072-970 Natal, RN, Brazil
         \and European Southern Observatory,
              Karl-Schwarzschild-Str. 2, D--85748 Garching b. M\"unchen, Germany
              }

   \date{Received 23 November 2005 / Accepted 15 January 2006}

\abstract
{}
{Analysis of the innermost regions of the carbon-rich star IRC+10216 and of 
  the outer layers of its circumstellar envelope have been 
performed in order to constrain its mass-loss history.}
{High dynamic range near infrared adaptive optics 
 and high angular resolution deep $V$-band images of its circumstellar envelope collected
 with VLT/NACO and VLT/FORS1 instruments have been analyzed.}
{Maps of the sub-arcsecond structures, or clumps, in the innermost regions are
 derived from the near-infrared observations. The morphology of these clumps is 
 found to strongly vary from  J- to L-band. 
 Their relative motion appears to be more complex than proposed in earlier works:
 they can be weakly accelerated, have a constant velocity,
 or even be motionless with respect to one another.
 From $V$-band imaging, a high spatial resolution map of the shell
 distribution in the outer layers of IRC+10216 is presented.  Shells are well
 resolved up to a distance of about 90$''$ to the core of the nebula and most of them
 appear to be composed of thinner elongated shells.
Finally, by combining the NACO and FORS1 images,
 a global view, showing both the extended layers
 and the bipolar core of the nebula together with the real size of the inner
 clumps is presented.}
{This study confirms the rather complex nature of the IRC+10216 circumstellar environment.
 In particular, the coexistence at different
 spatial scales of structures with very different
 morphologies (clumps, bipolarity and almost spherical external layers) is very puzzling.
 This confirms that the formation of AGB winds
 is far more complex than usually assumed in current models.}

 \keywords{stars: AGB and post-AGB -- stars: variables: general -- stars: individual: IRC+10216
           -- stars: mass-loss -- stars: circumstellar matter -- techniques: high angular resolution
           }

 \maketitle
%

\section{Introduction}

Low- and intermediate-mass stars lose a large amount of their
initial mass when they evolve along the Asymptotic Giant Branch (AGB)
and beyond.
During these mass-loss events, a huge circumstellar envelope (CSE) is formed.
\object{IRC+10216} is the best-known example of such evolved stars with an optically
thick CSE.
Indeed, its envelope almost completely absorbs
the central stellar photons in visible light and at shorter wavelengths.
This circumstellar environment has therefore been mostly studied
in the infrared and millimeter domains, spectral regions where the
envelope radiates itself and scatters the stellar light.

At very small scales (arcsec and below), a detailed picture of the IRC+10216
central regions has already been presented by several groups
(see~e.g.~Haniff \& Buscher~\cite{HanBu98}; Weigelt et al.~\cite{Wei98}, \cite{Wei02};
Tuthill et al.~\cite{Tut00}, \cite{Tut05}). The central core appears to be
composed of a 
series of clumps whose positions and luminosities vary
on time-scales of a few years. The complexity of the structures
detected has led to several hypotheses regarding
the precise location of the central star.

At much larger scales (up to arcmin),
Mauron \& Huggins~(\cite{Mau99}, \cite{Mau00}, MH99-00 hereafter) have shown that
the IRC+10216 CSE can also be studied in visible wavelength if enough
deep images are collected. 
The nebula brightness then results from galactic ambient light scattered
by its dust particles.
It is detected 
up to very large distances to the central star (up to about 6\,000 stellar
radii) and 
thus carries information about the mass-loss history during the
last few thousand years. 
MH99-00 have also shown that this fairly round circumstellar 
envelope is consistent with an
isotropic galactic radiation field and a spherically symmetric dust shell
(see also Mauron et al.~\cite{Mau03}).
However, on a better spatial resolution ($\sim 1$~arcsec), the envelope consists of a
series of 
discrete and nested multiple shells (or arclets) whose origin is still debated.
Although IRC+10216 is the only known AGB with such shells,
similar morphology has already been detected
around a dozen of planetary nebulae (PN)
and about six proto-planetary nebulae (PPN).
However, all these PN and PPN are bipolar,
contrary to what we observe for their progenitor (assuming that
IRC+10216 CSE properties are common for AGB stars). 
The cause and occurrence of the transition from a spherical
multiple-shell CSE to a bipolar one
is crucial for the understanding of the mass-loss phenomenon on the AGB
and the evolution of the material ejected into the interstellar
environment.

To date, no global view of the morphology
of the IRC+10216 CSE at different scales exists.
The aim of this work is to provide such a global description
by combining new high dynamic and high spatial resolution images of
its innermost regions collected with adaptive optics techniques
together with new deep images of its most external layers.
These observations are presented in Sect.~\ref{secobs}.
We analyze in Sect.~\ref{secNACO} the morphology of the innermost regions 
and their temporal variations.
Sect.~\ref{secFORS} is devoted to the analysis of the numerous shells
found in this envelope and to some of their properties.
We then discuss, in Sect.~\ref{secFORSNACO}, the coexistence of the different
morphologies found in the CSE of IRC+10216.
Finally, a conclusion is presented in Sect.~\ref{secConcl}.


\section{Observations and reductions \label{secobs}}

\subsection{NACO observations \label{secNACOobs}}

Infrared images of IRC+10216 were recovered from ESO Science Archive Facility.
They were obtained in November 2002 and March 2003, using the adaptive optics system NACO at the ESO/VLT Yepun telescope.
NACO is an association of the adaptive optics system NAOS (Rousset~\cite{Rou00}) and the spectro-imager CONICA (Lenzen~\cite{Len03}).

We have recovered observations of IRC+10216 obtained with the narrow-band filters
NB~$1.24$ (centered at $\lambda_c$~$=1.237$~$\mu\rm{m}$, $\Delta\lambda$~$=0.015$~$\mu\rm{m}$),
NB~$1.26$ ($\lambda_c$~$=1.257$~$\mu\rm{m}$, $\Delta\lambda$~$=0.014$~$\mu\rm{m}$),
NB~$1.64$ ($\lambda_c$~$=1.644$~$\mu\rm{m}$, $\Delta\lambda$~$=0.018$~$\mu\rm{m}$),
NB~$1.75$ ($\lambda_c$~$=1.748$~$\mu\rm{m}$, $\Delta\lambda$~$=0.026$~$\mu\rm{m}$),
NB~$2.17$ ($\lambda_c$~$=2.166$~$\mu\rm{m}$, $\Delta\lambda$~$=0.023$~$\mu\rm{m}$)
and the broad-band
filter L$'$ ($\lambda_c$~$=3.80$~$\mu\rm{m}$, $\Delta\lambda$~$=0.62$~$\mu\rm{m}$).
The pixel scale on CONICA was respectively 13.27~mas in the narrow-band filters and 27.15~mas in the L$'$ filter.
Observation conditions, as well as total on-source integration time for each filter, are summarized in Tab.~\ref{tabobs}.
Calibration files (flat fields and dark exposures) and observations of the PSF reference star HR~3550 were also recovered.
The Jitter technique was used in all observations.
The box size of the L$'$ broad-band image was about $7'' \times 7''$,
and the box sizes in the narrow-band images varied between about $4'' \times 4''$ and $7'' \times 7''$.
As shown in Tab.~\ref{tabobs},
the seeing conditions were variable, ranging between about $0.5''$ and $0.8''$.
The best dynamic ranges of the IRC+10216 final images (not deconvolved) were
around $7\,000$, $40\,000$, $6\,000$ and $90\,000$ AFU (Arbitrary Flux Units) for the J, H, K and L bands, respectively.
The noise level was found to be smaller than 30 AFU in all images.
In the PSF observations, the seeing varied between $0.5''$ and $0.6''$ and the air-masses between about $1.3$ and $1.4$.
The estimated FWHM of the PSF star was around 70~mas in the J and H bands, 80~mas in the K-band and 120~mas in the L-band.

Standard reduction procedures were applied using self-developed routines.
The raw images were sky subtracted, then divided by the flat-field and corrected from hot pixels.
In each filter, the images were cross-correlated and aligned by sub-pixel shifting, and then combined to produce the final images,
eliminating cosmic rays hits.
Finally, they were deconvolved with the PSF reference star.
We used the Richardson-Lucy algorithm (Richardson~\cite{Rich72}; Lucy~\cite{Lucy74}).
Since no PSF data for IRC+10216 were found in the November 22, 2002 observations, we have developed for that night pseudo-PSF images,
by analyzing and comparing the other IRC+10216 observations with their corresponding PSF data.
Constancy of prominent features present in deconvolved images showed that the PSF selection and the number of iterations
(25 typically) for the deconvolution process was performed carefully and
conservatively.
We have then summed the deconvolved
images in each band
(see Fig.~\ref{figJHKL}).
The highest dynamic range J-band image was obtained from the $1.24 \mu$m and $1.26 \mu$m images, which leads to about $14\,000$ AFU.
For the H-band, we have summed both $1.64 \mu$m and $1.75 \mu$m narrow-band images, obtaining a dynamic range of about $78\,000$ AFU.
The three $2.17 \mu$m narrow-band images were combined to build a K-band image
($\sim 16\,000$ AFU).
Finally, the L-band image has about $92\,000$ AFU.
These images have thus the best dynamical range ever published (see e.g. Tuthill et al.~\cite{Tut05}).
We note that, over the interval of about 4 months
between the first and last observations studied here,
no clear variations of the positions of the structures were found.
\begin{table}
 \caption{NACO observations log of IRC+10216.}
 \centering
 \begin{tabular}{ l c c c c r }
  \hline \hline
  \small
  Date      & Filter     & On-source          & Seeing  & Air-    & Dyn.          \\
            &            & exp. time          &         & mass    & range         \\
  (UT)      &            & (sec)              & ($''$)  &         & (AFU)         \\
  \hline
  22 Nov 02 & NB $1.64$  & 128                & $0.6$   & $1.5$   & $28\,700$     \\
            & NB $2.17$  & 120                & $0.6$   & $1.6$   & $6\,200$      \\
  16 Mar 03 & NB $1.26$  & 120                & $<0.5$  & $1.3$   & $6\,900$      \\
            & NB $1.64$  &  70                & $<0.5$  & $1.3$   & $8\,300$      \\
            & NB $2.17$  &  60                & $<0.5$  & $1.3$   & $4\,800$      \\
  18 Mar 03 & NB $1.24$  & 210                & $0.6$   & $1.3$   & $7\,200$      \\
            & NB $1.75$  & 103                & $0.7$   & $1.3$   & $41\,500$     \\
            & NB $2.17$  & 200                & $0.8$   & $1.3$   & $5\,200$      \\
            & L$'$       & 183                & $0.6$   & $1.3$   & $92\,500$     \\
  \hline
 \end{tabular}
 \label{tabobs}
\end{table}

\subsection{FORS1 observations \label{secFORSobs}}

The observations were collected with the VLT-Antu telescope,
equipped with the FORS1 focal reducer.
The detector is a 2048$\times$2048 thinned 24\,$\mu$m pixel Tektronix chip.
The field of view of individual images is $6.8' \times 6.8'$ with a pixel
size of 0.2~arcsec (see Appenzeller et al.~\cite{App98}).
All the exposures
were acquired in standard FORS1 service mode using a classical Bessel
$V$-band filter.
The available data consists of eight
10-min exposure and two 20-min exposure frames collected from
10 to 11 January, 2000, leading to a total observing time of 2~hours.
The selected exposures were taken in dark time under
very good seeing conditions and photometric sky. A few other 10-min
frames were indeed rejected
due to their moderate seeing conditions.
The mean airmass
was 1.3.
The telescope was shifted
by a few arcseconds between each individual image.

It was found that individual images reduced by
the standard ESO reduction pipeline (which includes standard corrections
such as bias subtraction, flat fielding, etc., see Hanuschik \& Amico~\cite{Han00})
were of rather poor quality.
We suspect this was due to the use of a corrupted flat field. Therefore,
a new reduction procedure was performed for all individual
exposures (removal of cosmic and aberrant pixels, flat-fielding with
a specific mean sky flat for each night, etc.).
All exposures taken on the
same night were then shifted and stacked. The final reduced image was built by
adding the summed exposures collected during the same night, taking
into account of their respective total exposure time.
It consists of 1900 $\times$ 1900 pixels corresponding to a total
field of view of
$6.3' \times 6.3'$. The resulting mean seeing, measured
from the brightness profile of individual stars, is found to be around
0.65$''$. The central core of IRC+10216 is measured with a S/N ratio
larger than 100 per pixel and the S/N of the envelope at 20$''$ from the
center is around 5-6 per pixel.
For a more detailed description of this reduction procedure
see Vandame~(\cite{Van02}),
and a preliminary presentation of this image can be found in de Laverny~(\cite{Lav03}).
Due to the wide field of this image, we have estimated a PSF
reference by using the median average of a set of suitable point-like stars, that
have been, firstly, background-subtracted, centered with
sub-pixel accuracy and normalized.
The final FORS1 image (see Sect.~\ref{secFORS}) has been deconvolved
using a Richardson-Lucy algorithm (Richardson~\cite{Rich72}; Lucy~\cite{Lucy74}).
This procedure has slightly improved the spatial resolution (to less than about 0.6$''$)
     and the S/N ratio (to about 7-8 per pixel at 20$''$ from the center).


\section{The CSE innermost regions \label{secNACO}}

   \begin{figure*}
   \centering
   \includegraphics[width=18cm]{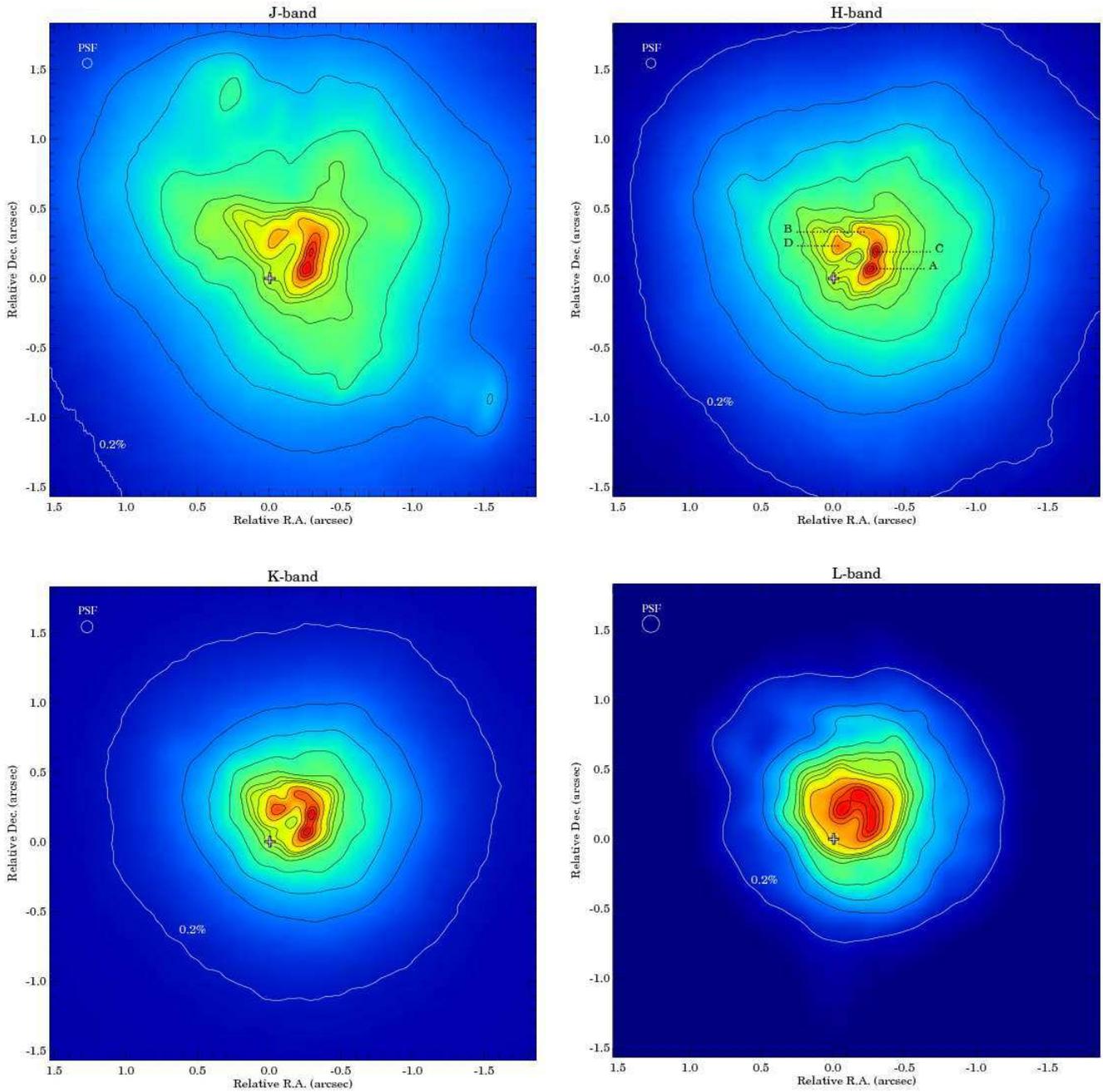}
   \caption{NACO JHKL images of IRC+10216.
            Contour levels are 80, 60, 40, 20, 10, 8, 6, 4, 2, 1, 0.5, and
            0.2\% of the peak surface brightness.
            North is up and East is left.
            The white cross at each image center represents the assumed central star
            position and its size is proportional to the error of 0.03'' as given by Murakawa et
            al.~\cite{Mur05}.
            Clumps A to D of the H-band follow the Haniff \& Buscher~(\cite{HanBu98}) clump nomenclatures.
            The resolution is about 70~mas in the J and H bands, 80 mas in K and 120~mas in L
            (represented by the circles at each image corner).
            }
            \label{figJHKL}
    \end{figure*}

The JHKL diffraction-limited images of IRC+10216 are displayed in
Fig.~\ref{figJHKL} 
with a log-scale for the brightness, so that details of the morphology at all flux levels can be seen.
The labels A to D shown in the H-band image 
indicate the features identified and labeled by Haniff \& Buscher~(\cite{HanBu98}).
The images have been centered at the central star location estimated by Murakawa et al.~(\cite{Mur05}).
These authors have performed a polarimetric study of IRC+10216 in H-band
which has independently provided a possible central star position, after series of contradictory hypothesis (see Weigelt et al.~\cite{Wei02} and Tuthill et al.~\cite{Tut05}).
Following Murakawa et al.~(\cite{Mur05}),
we have used clump A as reference to identify the central star position.
We note that their observations were made at the same epoch as the images presented here.

\subsection{Morphology at different wavelengths \label{secNACOcommon}}

The images exhibit a bright and inhomogeneous
structure which roughly looks like a
ring (with a diameter of approximately $0.5''$) 
composed of clumps (including clumps A to D) around an approximately circular depression.
The depression is located at about $(0.15,0.15)''$ from the image center and
has about 6\% of the intensity peak.
Clumps A and C are the brightest features.
Clump B appears as an elongated feature, at about the NE direction from clump C,
and clump D seems to be a more spread out feature.
In addition, there is a faint and almost spherical extended envelope
(from $\sim 0.5''$ to more than $1''$ from the image center), 
which also appears to have its center in the ring depression.
This depression could thus correspond to an apparent center of the images.
Regarding the central star, it is located in the fainter SE region of the mentioned ring.
Its position also coincides with a particular elongation in the ring brightness distribution,
well seen in the H-band image.
Finally, we have verified that the faintest structures seen in the J-band
are ghosts, probably due to the light reflected in the NACO optics.

On another hand, a clear difference between the JHKL images concerns the brightness variations of
clumps A to D with respect to the images peaks.
Clump A remains close to the intensity peak in all bands whereas
clumps B and D are brighter at larger wavelengths (from about 10\% and
20\%, respectively, of the intensity peak in J to about 100\% in L).
The brightness of clump C increases more slightly and is always brighter than 80\%.
We also note that the brightness difference between the four clumps
strongly decreases with increasing wavelength.
The SE region of the ring, close to the assumed location of the central star, remains faint,
varying from about 8\% to 20\% of the intensity peak from J to L.
Finally, the
extent of the external envelope seems to decrease with increasing wavelength.
Considering its limits at $0.2$\% of the image peak brightness,
we have calculated its mean extent as being about $4.8''$, $3.5''$, $2.7''$ and $2.0''$ in the J-, H-, K- and L-band, respectively.

The clump brightness variations as well as the extent of the envelope at different wavelengths
may reveal that we mostly detect, in K \& L, the emission of the heated dust, whereas at shorter wavelengths the scattered stellar emission becomes more important.
This is in agreement with models of the spectral energy distribution computed for IRC+10216
(see~e.g.~Mauron et al.~\cite{Mau03}).
In the K \& L bands, the dust emission is indeed $\sim100$ times
larger than the scattered stellar light
which becomes dominant below $\sim\,1\mu m$.
Therefore, we can deduce that most of the clumps seen close to the star in the L-band
have approximately the same temperature.
On the contrary, in the J-band optical depth effects
could explain the different brightness of the clumps.

\subsection{Temporal variations \label{secNACOt}}

Temporal changes of the IRC+10216 innermost regions have already been reported
by Tuthill et al.~(\cite{Tut00}), Weigelt et al.~(\cite{Wei02}) and
references therein.
Weigelt et al.~(\cite{Wei02}) have estimated
approximately linear displacements between clumps A-C and A-D,
and a possible acceleration of 5~mas~yr$^{-2}$ for the separation A-B.
Tuthill et al.~(\cite{Tut00}, \cite{Tut05})
have identified two sub-components in clump B: NE1 and NE2
close and far from clump C, hereafter referred as B$_1$ and B$_2$, respectively. They
 have proposed that clumps
B$_1$, B$_2$ and D move away from A, apparently with an uniform acceleration of 3.4~mas~yr$^{-2}$.

\begin{figure}
   \centering
   \includegraphics[width=8.9cm]{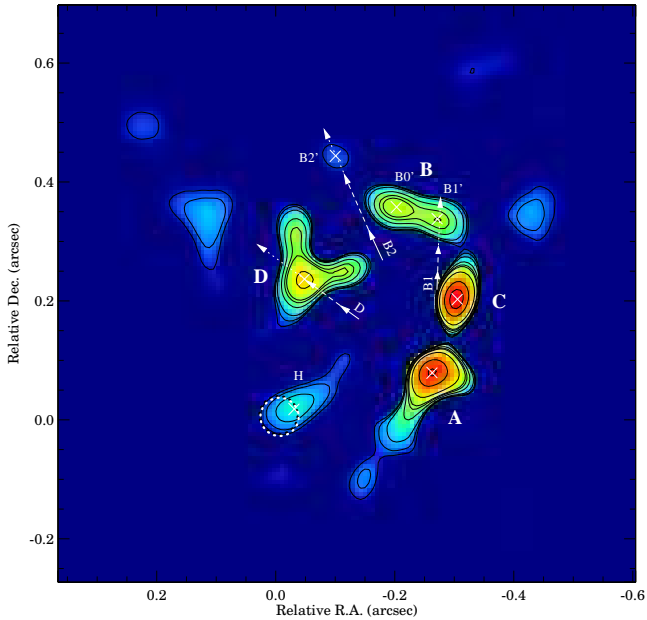}
   \caption{H-band map of IRC+10216, where
            only the highest spatial frequencies of the brightness
            in the Fourier space have been kept.
            The contour levels are 80, 50, 20, 10, 8, 5, 2, 1 and 0.5\%
            (this minimum level being the estimated noise).
            The dotted circle is the assumed star position, as in Fig.~1.
            Main clumps are indicated by the labels A to D,
            and some sub-features by the smaller labels.
            The positions of these clumps are shown by the white crosses.
            Estimated trajectories (from the results of Tuthill et al.~\cite{Tut00})
            for some clumps with respect to clump A
            are also shown (see text for details).
            }
            \label{fighif}
    \end{figure}

We have also applied on the
H-band\footnote{We have selected the H-band image
(instead of the K-band most commonly analyzed)
because of its better spatial resolution and considerably better dynamic range.
We have verified that the detected clumps and their estimated positions
are similar to those found in the K-band.}
image a Fourier filtering procedure in order to remove
the structures of lower spatial frequencies (see Fig.~\ref{fighif}).
We have then identified new features,
in particular, the sub-features B$_0'$, B$_1'$ and B$_2'$, in clump B.
We found a feature, not identified in previous works, 
close to the assumed star position (labeled H).
We note that the star position assumed in this work could still be discussed
and its connection with clump H is very unclear.
This clump could be, for instance, a dust cloud just passing between the star and the observer.
Future observations are needed to study the evolution
of this clump with respect to the central star position.

Fig.~\ref{fighif} shows estimated apparent trajectories for the previously detected clumps
B$_1$, B$_2$ and D, with respect to A.
These estimations were made by assuming that the clumps move away from A,
as proposed by Tuthill et al.~(\cite{Tut00}), and by taking into account of their spatial separations.
The solid arrows represent the displacement of these clumps during the
interval time of their observations, i.e. from 1997 January to 1999 April.
The dashed arrows show a prediction for the clumps displacements up to 2003 March,
by assuming the averaged velocity of Tuthill et al.~(\cite{Tut00}).
The dotted arrows represent an alternative prediction by assuming the acceleration law
proposed by Tuthill et al.~(\cite{Tut00}).
The error margins are about 10~mas for the dashed arrows and 30~mas for the dotted ones.
We can see that the previous clumps B$_1$ and B$_2$ are most probably the current sub-features
B$_1'$ and B$_2'$, respectively.
They are currently separated by 258~$\pm 20$, 394~$\pm 20$ and 261~$\pm 20$ mas from A.
B$_1$ and B$_2$ thus appear to be less accelerated than expected.
At the same time, clump D appears to have moved with a constant
velocity.
>From the clumps separations given by Weigelt et al.~(\cite{Wei02}),
we have also verified that clump C
(currently located at 131~$\pm 20$~mas from A)
appears to be approximately motionless. Note that choosing clump A as
reference could give the illusion that the clumps escape from it.
The clumps motions are therefore
not compatible with the uniform acceleration law
proposed by previous studies,
although some accelerations may exist for clumps B$_1$ and B$_2$.
New high angular resolution observations are needed
to disentangle the three-dimensional morphology of the innermost environment
of IRC+10216 and to study the temporal variations of these clumps.


\section{External layers of IRC+10216 \label{secFORS}}

   \begin{figure}
   \centering
   \includegraphics[width=8.9cm]{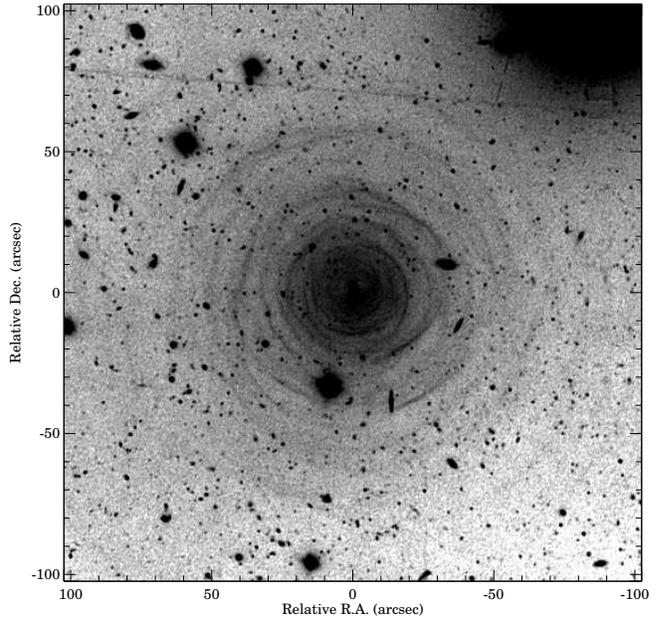}
   \caption{FORS1 deconvolved V-band image of IRC+10216.
            North is up and East is left.}
            \label{figFORS}
   \end{figure}

Fig.~\ref{figFORS} shows the deconvolved wide-field V-band image of IRC+10216.
As already shown by MH99-00,
we see an extended halo composed of thin and irregular multiple-shells.
They appear to be non-concentric and azimuthally incomplete.
The CSE is
seen due to external illumination by the ambient Galactic light, scattered by the dust.
Since these photons can easily
penetrate into the CSE (their optical depth from the outside
towards the center being very low), the incomplete shells do reveal
lower densities in some parts of the CSE. The shell discontinuities can
obviously not be caused by some shadowing effects due to more external material.

\subsection{Structure of the external layers \label{secFORSimages} }

\begin{figure}
   \centering
   \includegraphics[width=8.9cm]{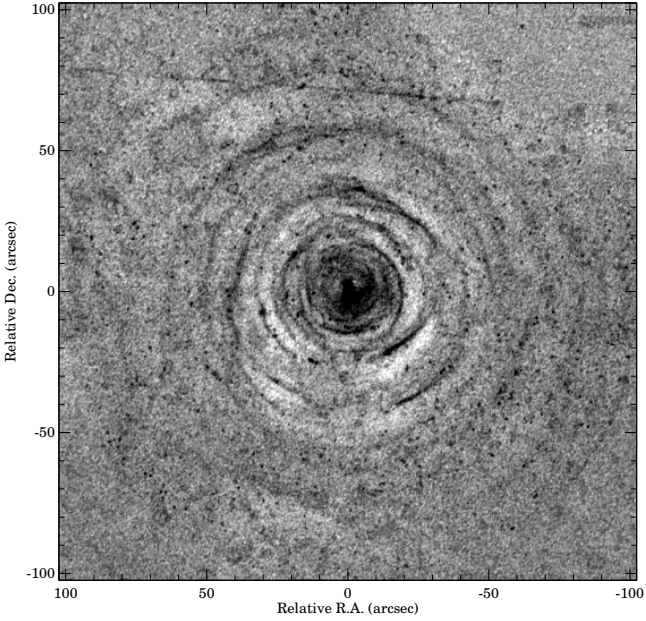}
   \caption{Deconvolved V-band image after subtraction of the halo of the CSE
            and removing of most stars and galaxies.}
            \label{figFORSclean}
    \end{figure}

\begin{figure}
   \centering
   \includegraphics[width=8.9cm]{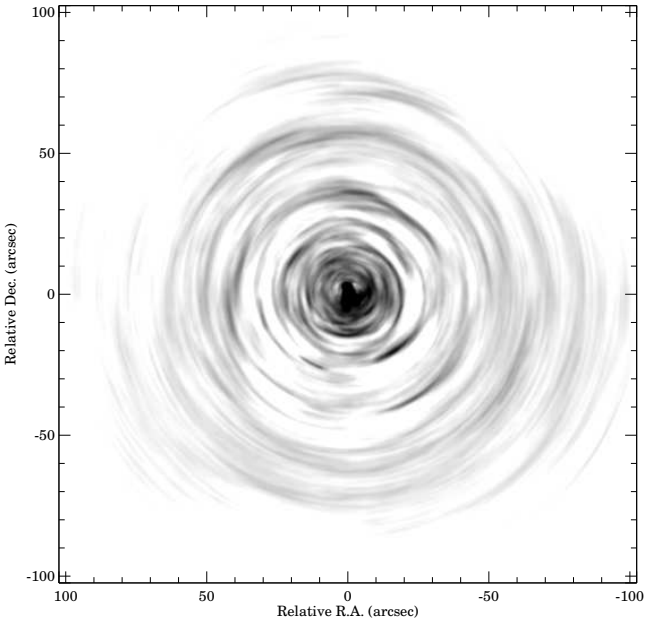}
   \caption{Schematic map of the shells surrounding IRC+10216 (see text for details).}
            \label{figFORSrings}
    \end{figure}

To emphasize the shell morphology, we have removed the central extended halo
by applying the same Fourier filtering procedure as for the NACO images.
We have also removed several sources (stars or galaxies) by selecting those
with observed intensities larger than a prefixed
 threshold. The source pixels were replaced by averaged values taking
 into account the local background and the noise level.
The resulting image is shown in Fig.~\ref{figFORSclean}.
We have then applied an azimuthal smoothing of 20$^{\rm o}$
around the center. Although this decreases the spatial resolution in the azimuthal direction,
the resulting map (Fig.~\ref{figFORSrings}) shows a clear visualization of the
shells,
and gives a more realistic and complete pattern than that presented in previous works.

The shell distribution seen in Figs.~\ref{figFORSclean}~\&~\ref{figFORSrings}
are similar to those reported before.
For instance,
the three faint and apparently thick shells at North,
located at about 30--40$''$, 50--60$''$ and 70--80$''$ from the center,
can also be seen in the CFHT image (MH99-00). However,
these shells, being better resolved in the FORS1 images,
appear to be composed of a complex sub-distribution of thinner ones.
Similar thin shells located very close each other are well seen everywhere in
the CSE. 
For instance, the shells labeled $e$ and $f$ by MH00
(located at distances to the center of about 55$''$ and 58$''$,
between 347--20$^{\rm o}$ and 23--53$^{\rm o}$, respectively)
are clearly mergers of complex thin shell distributions.
Another prominent shell located to the S, at about 15$''$ from the center,
joins series of slightly less prominent thinner ones
distributed toward SE, between 10--30$''$ from center.
Moreover, even in the more internal regions, several thin shells seem to
merge in thicker ones between about 4$''$ and 20$''$.
The whole CSE thus appears to be composed of a complex of several thin irregular
shells that could be identified as thicker ones in less resolved images.
Finally, we note that the separation between apparently thick shells varies a lot
with respect to the radial direction.

On another hand, a smooth azimuthally radial profile derived by computing the mean of all the
pixels found in annuli 0.9$''$ thick (see MH99-00 for more details)
confirms that the dust is detected up to about 200$''$ 
(about 5\,800 stellar radii).
That corresponds to material ejected about 8\,000 years ago
(assuming an escape velocity of 14~km~s$^{-1}$ and a distance of 120~pc), i.e.
an important fraction of an interpulse on the AGB.
Actually, we do not see any edge for the dusty envelope
and we are limited by the detector size.

\subsection{Thickness of the shells \label{secFORSthick}}

   \begin{figure*}
   \centering
   \includegraphics[width=17.8cm]{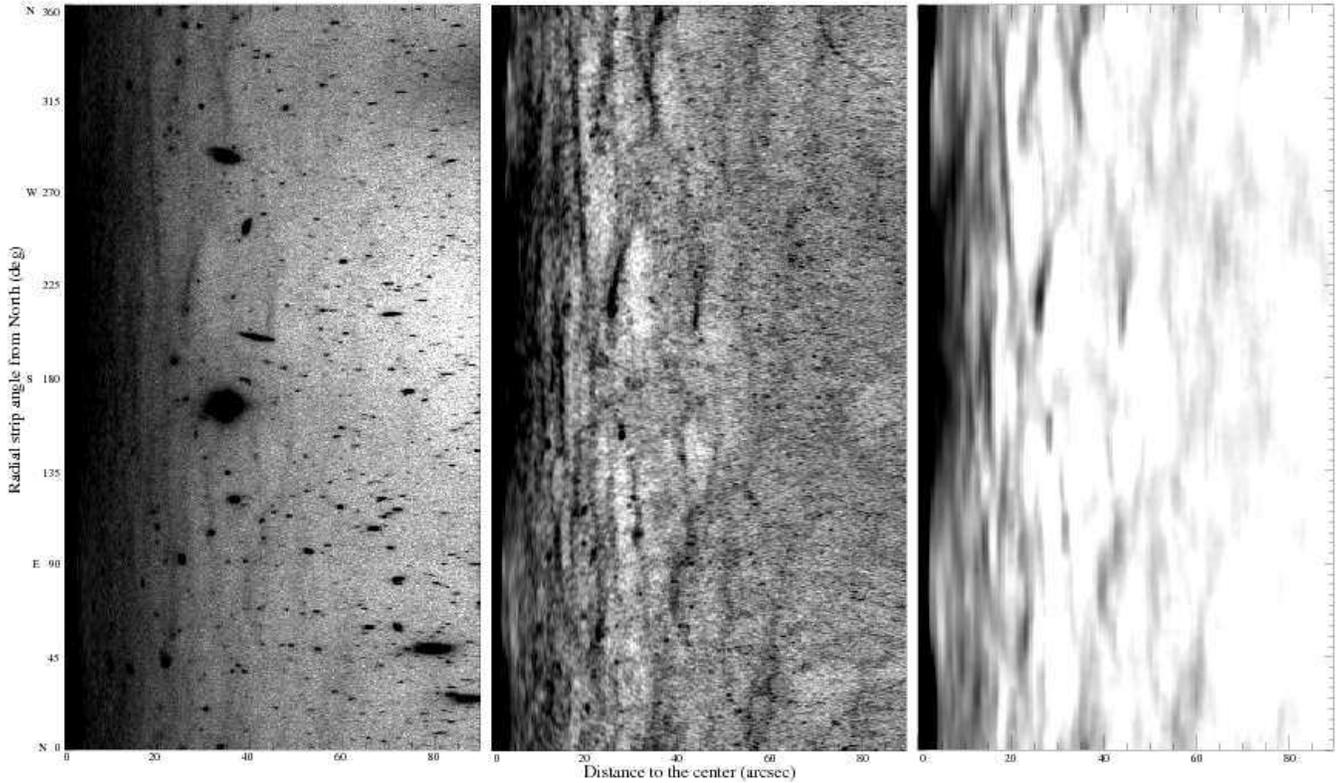}
   \caption{Map of the shells transformed from polar coordinates to a Cartesian representation.
            Each horizontal section of these maps represents a narrow radial strip of the FORS1 images,
            the angles of the strips being with respect to North.
            Maps to the left, center and right were derived from
            Figs.~\ref{figFORS}, \ref{figFORSclean} \& \ref{figFORSrings}, respectively.
            }
            \label{figFORSmap}
    \end{figure*}

MH00 have proposed
that the shells thickness increases with increasing the distance to the center, 
in agreement with the natural expansion of the envelope.
To analyze the shell profiles and to verify their finding, we propose here a
new and more accurate method,
by taking into account the non-concentric nature of the shells  and
the possibility that an apparently thick shell may be resolved into 
several thinner ones.

Fig.~\ref{figFORSmap} shows the CSE morphology in a map of
narrow\footnote{Radial sections with thickness of 1 pixel
                and a rotation step of 0.2$^{\rm o}$.
                Interpolations between the original image pixels were applied for each step.}
radial strips. From this diagram, we clearly see the non-concentric nature of the shells
as well as the complexity of their spatial distribution.
For instance,
the long and thin shell
located at distances to the center of about 15--20$''$, between 220$^{\rm o}$ and 340$^{\rm o}$
from North, has an inclination of about 4$^{\rm o}$ in this diagram
with respect to the vertical axis.
The shell located at distances to the center of about 30--35$''$, between 300$^{\rm o}$ and 360$^{\rm o}$ from North,
has an inclination of about 17$^{\rm o}$.
At the same time, there are shells with opposite orientations,
as those found at distances to the center of about 25--30$''$ and 45$''$,
from 210$^{\rm o}$ to 260$^{\rm o}$, and from 200$^{\rm o}$ to 230$^{\rm o}$, respectively,
which make angles of about $-16^{\rm o}$ and $-2^{\rm o}$ with respect to the vertical axis.

\begin{figure}
   \centering
   \includegraphics[width=8.9cm]{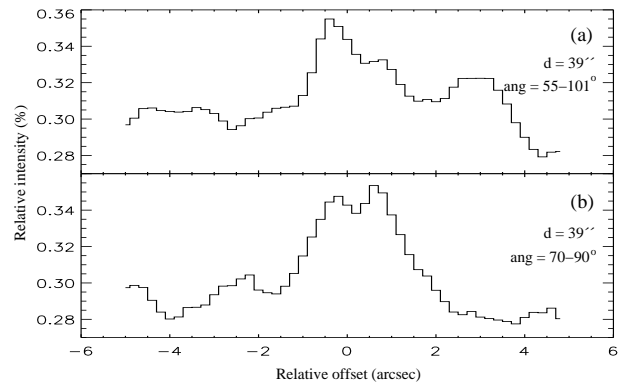}
   \caption{Comparison of the profile of the shell $d$ estimated as in MH00
   (top panel) with its profile derived by our more complex method (bottom
   panel, see text for details).
            The distances to the center, $d$, of the shells
            and the ranges of their azimuthal angles from north, {\it ang},
            are given.
            The relative intensity is with respect to the central peak brightness of the original image.
            }
            \label{figFORSprofd}
    \end{figure}

\begin{figure}
   \centering
   \includegraphics[width=8.9cm]{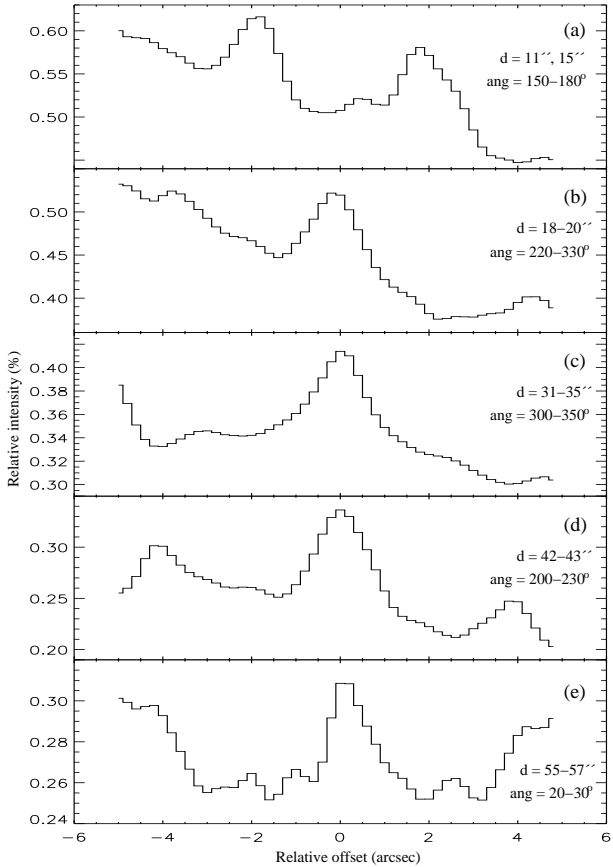}
   \caption{Profiles of some shells considering their non-concentric nature
            (see text for details).
            There are two shells in the panel (a) and one in the others.
            The given parameters follow the same definitions as in Fig.~\ref{figFORSprofd}.
            The ranges in the shell distances to the center, $d$,
            are due to their non-concentric nature.
            }
            \label{figFORSprof}
    \end{figure}

\begin{figure}
   \centering
   \includegraphics[width=8.9cm]{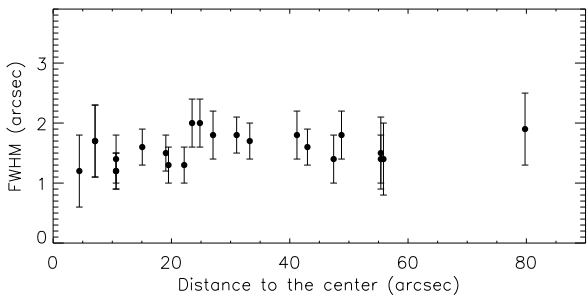}
   \caption{Relation between the thickness of the shells and their distance to the center.}
            \label{figFORSgrprof}
    \end{figure}

Regarding the profile of the thickest shells, we
have carefully analyzed the shell labeled $d$ in
MH00.
Fig.~\ref{figFORSprofd}(a) shows its profile by applying the same method as those authors.
Fig.~\ref{figFORSprofd}(b) shows the profile of the same shell estimated from 
the more restricted region
located at a distance to the center of about 39$''$, between 70--90$^{\rm o}$,
where it appears more regular. We have
then integrated
profiles perpendicular to its direction and subtracted the extended halo contribution.
The estimated FWHM of the profile~(a) is about 3.0$''$ (as in MH00), whereas it is about
2.6$''$ for the profile~(b).
The error margins are around 0.4$''$.
Although both estimates are in agreement within the error bars, 
a deeper analysis of this shell reveals that
even our profile~(b) could be widened due to a merging of two thinner ones.
Indeed, the profiles shown in Fig.~\ref{figFORSprofd} have two close peaks at offsets of about $\pm 0.5''$,
which leads us to suspect that there are two thin shells close together in this
region and, hence, not well spatially resolved.
We note that this shell was carefully analyzed,
this pair of peaks being identified in every derived profiles.
If we decompose the profile~(b) in two close shells,
their estimated FWHM are about 1.8$'' \pm 0.4''$.
In consequence, we have derived several shell profiles
by identifying, as above, well resolved thin shells at different distances to the center,
(see Fig.~\ref{figFORSprof}).
The profiles (a) to (d) have good S/N ratios.
The profile (e) having a worse S/N ratio
is actually a thin feature composing an apparently thicker shell
which was also detected by MH00.
The FWHM of the two features in the profile (a)
and the other four features in the profiles (b) to (e) are, respectively, about:
1.2$''$, 1.6$''$, 1.5$''$, 1.7$''$, 1.6$''$ and 1.4$''$.
The error margins are around 0.4$''$.
In a more general way,
we have estimated the FWHM of 23 shells at radial distances from 4$''$ to
80$''$ (see Fig.~\ref{figFORSgrprof}).
We found a mean FWHM value of 1.6$''$, with a standard deviation of 0.3$''$.
The minimum FWHM value of 1.2$''$ is
found for the innermost analyzed shell, located at about 4$''$ from center 
between 40--100$^{\rm o}$,
and also for two shells located at about 9$''$ and 11$''$ from the center,
between 150--180$^{\rm o}$ and 205--255$^{\rm o}$, respectively.
The maximum FWHM value of 2.0$''$ is found for the shells located at about 24$''$ and 25$''$,
between 105--135$^{\rm o}$ and 60--85$^{\rm o}$, respectively.
The furthest analyzed shell, located at about 80$''$ from center, between 30--45$^{\rm o}$,
has a FWHM~$\simeq 1.9''$.
The error bars vary between 0.3$''$ and 0.6$''$.
Thin shells
are thus detected in the whole envelope and even far from center.
We therefore cannot derive a clear increasing relation
between the shell thickness and the distance to the center such as that proposed by MH00.
However, the shells found rather close from center could be resolved into even thinner
ones. We therefore could have overestimated their thickness.
In consequence, either the slope of the thickness variation with distance
proposed by MH00 could still be valid but with very thin shells close to the
star,
or the shell thickness increases much less than that estimated by these authors.


\section{Global view of IRC+10216 \label{secFORSNACO}}

In order to better understand the possible links existing
between the almost spherical shells and the inner clumps,
we describe here the morphology of the inner CSE
from the FORS1 image together with the NACO data.

\begin{figure*}
   \centering
   \includegraphics[width=8.9cm]{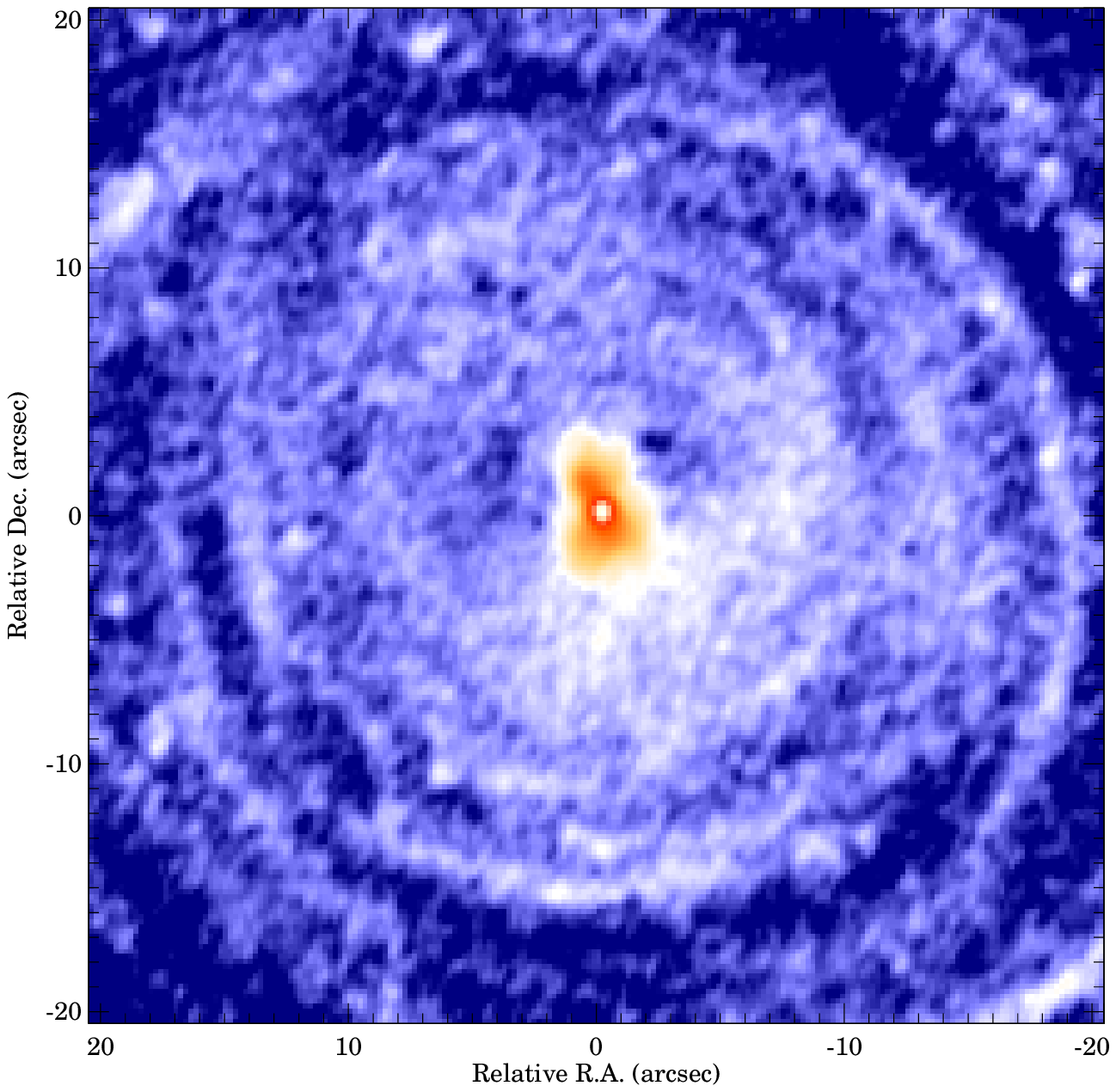}
   \includegraphics[width=8.9cm]{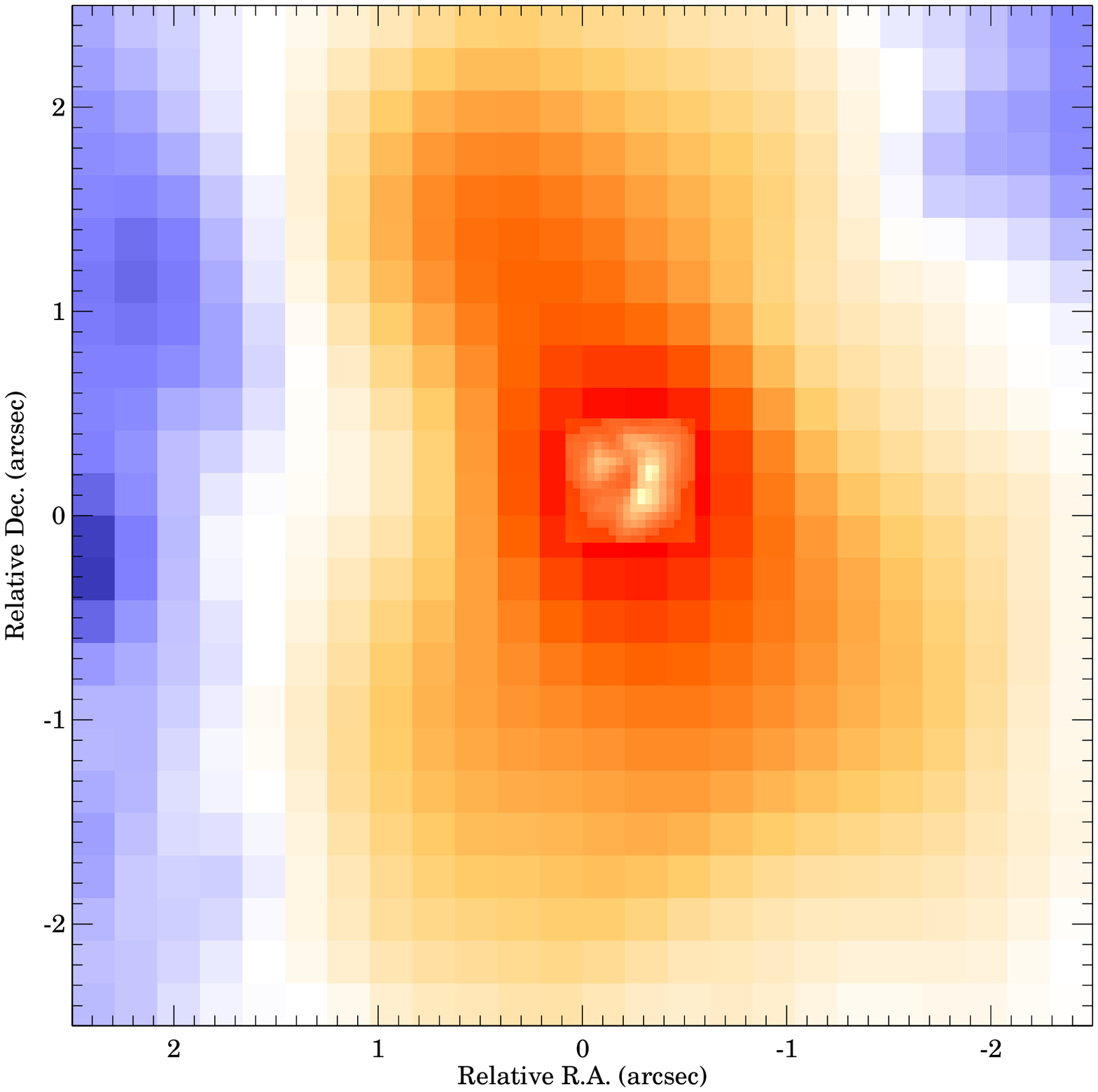}
   \caption{View of the IRC+10216 CSE inner morphology.
            Left panel shows the closest structures around the center detected in V-band.
            Right panel shows the V-band image core,
            on which has been superposed the NACO H-band image.
            The overlapping has been done by assuming that the NACO and FORS peak brightness
            are found at the same location.}
            \label{figforsnaco}
    \end{figure*}

Fig.~\ref{figforsnaco} (left panel) shows a closer view of Fig.~\ref{figFORSclean}.
MH99-00 have detected three structures suspected to be shells
in regions within about 3.1$''$ from the center, whereas 
no such shells are found in our data,
possibly because the HST data have a better spatial resolution,
despite their lower S/N ratio.
The closest identifiable structures
are located between $\sim$3$''$ and 16$''$ from center.
Regarding the core of the nebula, it appears clearly asymmetric.
Two dominant lobes
much brighter than the rest of the envelope lie around the center,
making together a direction of about 22$^{\rm o}$~$\pm 2^{\rm o}$ with respect to North.
The southern lobe being 40\% brighter than the northern one.
These features likely result from scattered stellar photons in contrast
to the reflected galactic light seen elsewhere in the envelope.
Such bipolar morphology
could be an indication that scattering is more efficient in the polar direction.
It could be roughly reproduced by a simple model
of scattering dust grains in a non-spherical dusty envelope, with evacuated 
polar regions, around the star, the system being tilted away from the observer
(see e.g.~Men'shchikov et al.~\cite{Men01}). We however note that the main
shape of the bipolar nebula slightly differs from the one reported by MH00
from their HST image,
possibly due to the different spatial resolution.

Fig.~\ref{figforsnaco} (right panel)
gives a representation of the IRC+10216 core,
by superposing the NACO and FORS images.
The two images were arbitrarily positioned by coinciding their intensity peaks.
We are conscious that this assumption may be crude
since the V and IR images result from very different physical processes.
However, this composite image represents
for the first time both the extended layers and the bipolar core together
with the real size of the inner clumps, and puts forward the difficulty of 
finding a link between such
small and large scale morphologies.
Firstly, evidence of clumps far from the center was not found
by Huggins \& Mauron~(\cite{HuMau02})
in a previous analysis of the same FORS1 image.
Secondly, shells cannot be
identified in the NACO images.
We note that, although the region composed of clumps A to D roughly looks like a ring,
its center (the depression feature) is not compatible with
the star position estimated by Murakawa et al.~(\cite{Mur05}).
Finally, the bipolar structure detected in V-band is also not clearly identified in the near-IR. 


\section{Conclusion \label{secConcl} }

We have described in this work very high quality images of the CSE
of IRC+10216, from its most inner regions to the most external ones.
In the central arcsec scale of the JHKL images,
 sub-arcsec structures (or clumps) identified
by other authors have been recovered about 4 years later. We have also derived a map of the 
brightest clumps found close to the core of IRC+10216.
The morphology of these clumps varies strongly with increasing wavelengths
and we propose  that the closest structures have about the same
temperature.
Furthermore, by analyzing their  
apparent relative motion, we cannot confirm the uniform outflow acceleration
previously proposed since only two bright clumps appear to be accelerated (but at a smaller
rate than that already estimated), whereas others clumps could have a constant velocity
or even no relative motion.
At much larger spatial scales (up to a few arcmin), we present a new
map of the non-spherical incomplete shells characterizing the CSE
of IRC+10216. Owing to the high spatial resolution of our image, most of the
thicker shells actually appear to be composed of thinner elongated ones.
Their thicknesses appear rather uniformly distributed between about 1$''$ and 2$''$,
regardless of the distance to the center.
Finally, we have combined the
NACO and FORS images in order to provide a more global view of this CSE
and to compare the typical size of the clumps
found very close to the center with the bipolar nebula and with the much
more external shells.

This study has confirmed the very complex nature of the IRC+10216 envelope
with asymmetries already present on the AGB.
Neither the morphology at different spatial scales nor the motions
detected very close to the center can be satisfactory explained
by current models on the mass-loss mechanisms in AGB stars
and their typical time-scales.
For instance, Sandin \& H\"ofner~(\cite{San04} and references therein)
have predicted shell density distributions
not compatible with those observed around IRC+10216
(see also on the same topic Meijerink et al.,~\cite{Mei03}).
Another scenario for the formation of the shells
in a spherically symmetric stellar wind
has been explored by Soker~(\cite{Sok00}, \cite{Sok02}).
He proposed that these shells
could be connected to cool magnetic spots on the stellar surface. 
If these spots are more concentrated
near the equator, the mass-loss geometry could deviate from sphericity
and thus favor the formation of shell-like features and/or clumps.

Moreover, it is interesting to note that the very complex structures
found around IRC+10216 may affect the chemical composition of its envelope.
For instance, the clumps detected very close to the central core may favor,
by their thermodynamical properties, 
the formation of the graphite observed in presolar dust grains
(Bernatowicz et al.~\cite{Ber05}).
Furthermore, the presence of high density shells in the photochemically
active regions could change the molecular distribution in the envelope
by blocking external UV photons
(see e.g.~MH00; Brown \& Millar \cite{Bro03}).
Then, high contrast shells of
complex molecules may be formed more easily, as confirmed by some millimeter observations
(see for instance, HCO$^+$, C$_2$H, C$_4$H and HC$_5$N maps by
Gu\'elin et al.,~\cite{Gue00},
and CO maps by Fong et al.,~\cite{Fon03}).

Finally, future high spatial resolution images of this CSE are still mandatory
in order to better understand the motions of its clumps (and in particular the
clump H superposed on the assumed central star), their
formation/fading, the central star position and the possible
evolution of the external shells as their three-dimensional
morphology.

\begin{acknowledgements}

We thank N. Mauron for fruitful and stimulating discussions over all these
years
and his comments on the manuscript,
and D.~O'Brien for proofreading it.
The Brazilian agencies CAPES and CNPq are thanked for financial support.
P.~de~Laverny acknowledges support from the CNRS/INSU ({\it Actions 
Th\'ematiques Innovantes})
and MESR {\it (Jeunes Chercheurs)}.

\end{acknowledgements}

\end{document}